\documentclass[aps,prl,twocolumn,showpacs]{revtex4-1}

\usepackage{graphicx}
\usepackage{amsmath}
\usepackage{amssymb}
\usepackage{color}

\begin{document}

\title{Topological aberration of optical vortex beams and singularimetry of dielectric interfaces}

\author{Mark R.~Dennis$^1$ and J\"org B.~G\"otte$^2$}
\affiliation{\\
$^1$H. H. Wills Physics Laboratory, University of Bristol, Tyndall Avenue, Bristol BS8 1TL, UK\\
$^2$Max-Planck-Institute for the Physics of Complex Systems, N\"othnitzer Str. 38, 01187 Dresden, Germany}

\date{\today}

\begin{abstract}
The splitting of a high-order optical vortex into a constellation of unit vortices, upon total reflection, is described and analyzed.
The vortex constellation generalizes, in a local sense, the familiar longitudinal Goos-H\"anchen and transverse Imbert-Federov shifts of the centroid of a reflected optical beam.
The centroid shift is related to the centre of the constellation, whose geometry otherwise depends on higher-order terms in an expansion of the reflection matrix.
We present an approximation of the field around the constellation of increasing order as an Appell sequence of complex polynomials whose roots are the vortices, and explain the results by an analogy with the theory of optical aberration.
\end{abstract}



\maketitle

Understanding and manipulating the spatial structure of light beams is a fundamental theme of modern optics.
Beyond the simple structure of plane waves, beams with inhomogeneous complex amplitude can carry quantized orbital angular momentum (OAM) \cite{Allen+:PRA45:1992}, which has many applications, including free-space classical and quantum communication \cite{Gibson+:OE12:2004,Paterson:PRL94:2005,Pors+:OE19:2011}, superresolved microscopy \cite{Maurer+:LPR5:2011} and extra-solar planet detection \cite{Foo+:OL30:2005,Serabyn+:Nature464:2010}

Free space optical modes carrying quantized OAM have optical vortices (phase singularities) on their axis \cite{NyeBerry:PRSLA336:1974,Dennis+:PO53:2009}: they have the amplitude structure $r^{|\ell|}\exp(i \ell \phi)$ close to the beam axis, for $\ell$ an integer: around the axis, as the intensity becomes zero, the phase gradient becomes infinitely large.
The core region close to the axis of such beams is therefore exceptionally sensitive to any imperfections in optical components along the path of the beam.
Any disruption to the pure azimuthal phase structure around the axis of the beam breaks the axial, strength-$\ell$ vortex into a constellation of $|\ell|$ unit strength zeros \cite{Freund:OC159:1999}. 
We call this effect \emph{topological aberration}, as the effect of aberration disrupts the topology of the simple optical mode.

Here, we describe in detail the topological aberration which an OAM-carrying beam experiences under dielectric reflection by an oblique angle.
We will see that the resulting constellation of zeros from the original strength-$\ell$ vortex depends on an aberration-like analysis of the complex reflection coefficient in Fourier space, which is sensitive to terms up to order $|\ell|$ in its Taylor expansion: it is therefore possible, using the spatial distribution of phase singularities, to determine higher orders of the effective reflection-induced aberration of the beam. 
Furthermore, this phenomenon will affect any vortex beam reflected by a mirror, prism or beam-splitter; as measurement techniques become more sophisticated, appreciation of this universal azimuthal symmetry breaking will become increasingly important.
Although the fact that a high-order vortex breaks apart upon perturbation has been long understood, our main result here is a formula for aberrated constellation of vortices.

Tracking the change of position of optical vortices on reflection echoes the study of optical beam shifts, where the centroid of a homogeneously-polarized beam is shifted by an amount proportional to the optical wavelength, either in the plane of incidence (Goos-H\"anchen shift \cite{GoosHaenchen:AndP436:1947}), transverse to it (Imbert-Federov shift \cite{Fedorov:DAN105:1955,Imbert:PRD5:1972}), or some combination of the two.
The net shift requires the initial transverse polarization to be homogeneous, and that the beam be narrow and axisymmetric in intensity, but is otherwise independent of the original amplitude profile.
In recent years, there has been an explosion of interest in various types of beam shift, including both fundamental insight and possible applications \cite{*[{see }] [{ and references therein}] BliokhAiello:CUP:2012}.

It is simple to see the connection between beam shifts and aberration: for total reflection in the plane of incidence (p-polarization) or perpendicular to it (s-polarization), the reflection coefficient \cite{Jackson:JohnWiley:1998} is a simple `tilt' \cite{BornWolf:CUP:2003} to first order; as the reflection coefficient acts on direction, in real space the reflected beam is translated. 
In partial reflection when the modulus of the reflection coefficients vary, there are also angular shifts of the beam's propagation \cite{Merano+:NatPhot3:2009}.

The notion of using the displacement of an on-axis zero to study beam shifts goes back to Hans Wolter in 1949 \cite{Wolter:ZN5a:1950b} who observed that the position of a zero is a precise marker depending on local information, rather than the beam centroid, which depends on the global distribution of intensity.
Our approach may be thought of as generalizing Wolter's approach to any incident polarization and vortex order, all of which may be further extended to other aberrations of vortex beams, including refraction.

Optical vortices are objects of scalar optics, and hence it is natural in our framework to consider a \emph{polarized component} of the reflected beam; beam shifts for polarized components are analogous to quantum `weak values' of the reflection operator \cite{DennisGoette:NJP:2012}, which have shifts in both real and Fourier space.
The spatial shift to the centroid of vortex-carrying beams involves both `spatial' and `angular' parts of phase-flat beams \cite{Fedoseyev:OC193:2001,BliokhDesyatnikov:PRA79:2009,Merano+:PRA82:2010}; we will see that this general shift is simply the first term, analogous to tilt, in an aberration expansion of the polarized component's reflection matrix.
In our calculation, higher aberration orders of the reflection matrix are realized as coefficients of a complex polynomial approximating the low amplitude in the core of the reflected beam: the vortex positions are related to these aberration terms by the fundamental theorem of algebra.

\begin{figure}
\begin{center}
\includegraphics[width=0.49\textwidth]{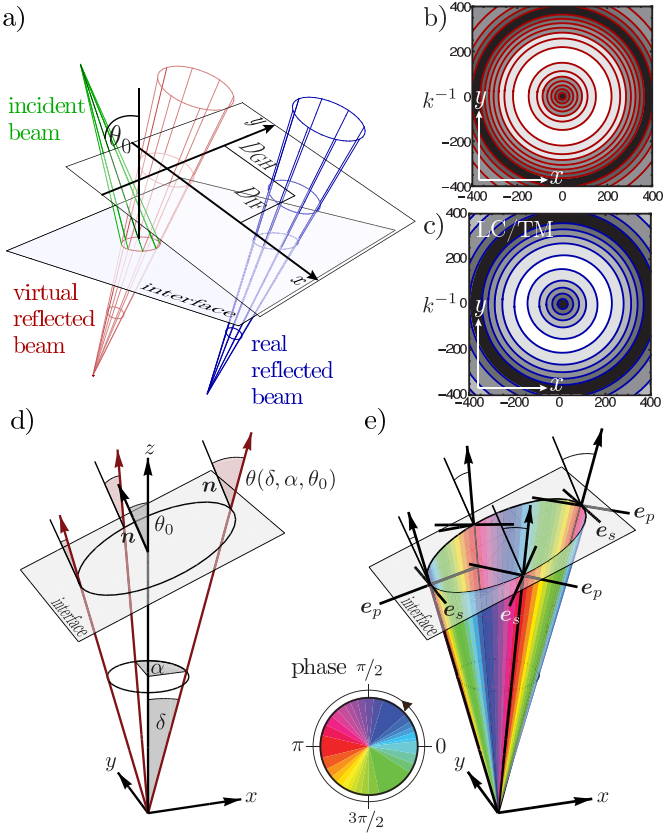}
\end{center}
\caption{\label{fig1} (Colour online)
   Incident and reflected beam geometry.
   a) Schematic of the relation between interface plane and our coordinate frame centred at the virtual reflected beam (red), with intensity profile given in b). 
   The green cone depicts a section of constant $d$ of the incident beam, the blue cone is the shifted, real reflected beam, with intensity as in c) for $n=2/3, \theta_0=44^\circ$.
   d) Schematic of the beam coordinates. 
   The local plane and angle of incidence depend both on $\delta$ and $\alpha$. 
   e)  Beam coordinates for a $\ell = -4$ vortex beam (phase indicated by colour wheel) and local resolution into $\mathbf{e}_s$ and $\mathbf{e}_p$.
   } 
\end{figure}

The basic beam geometry is demonstrated in Fig.~\ref{fig1}, choosing a natural coordinate system based on the construct of a virtual reflected beam \cite{GoetteDennis:NJP:2012}, which is obtained by specular reflection of every plane wave in the incident beam.
The central propagation direction is $(0, 0, 1),$ and we will make much of spherical angles about this direction: azimuth $\alpha$ and polar angle $\delta.$  
Since reflection flips the sign of $\ell,$ the reflected beam is $\sigma(\delta) \exp(-\mathrm{i}\ell \alpha),$ with real $\sigma(\delta)$ tightly centred around $\delta=0,$ such that the second moment $\sigma_2= \int_0 d \delta | \sigma(\delta)|^2 \delta^3 \ll 1$ (where the extra $\delta \approx \sin \delta$ is the jacobian in spherical polars).
The upper limit of this integral is larger than the width of the spectrum. 
In these coordinates, the normal to the incidence plane has direction $(-\sin \theta_0, 0, \cos \theta_0),$ so, for each component of the spectrum labelled by $\delta, \alpha$, the incidence angle $\theta$ is given by $\cos \theta = \cos \delta \cos \theta_0 - \cos \alpha \sin \delta \sin \theta_0.$ 
This angular dependence appears in the reflection coefficients $r_s(\theta)$ and $r_p(\theta)$ \cite{BornWolf:CUP:2003}, which are defined with respect to the local plane of incidence for every plane wave within the spectrum (see  Fig.~\ref{fig1}d).
The indices $s$ and $p$ thereby refer to the directions orthogonal $(s)$ and parallel $(p)$  to the local plane of incidence (see Fig.~\ref{fig1}e) and the general reflection matrix $\mathbf{R} = r_s \mathbf{P}_s - r_p \mathbf{P}_p$ consists of the projectors $\mathbf{P}_{j} = \boldsymbol{e}_j \otimes \boldsymbol{e}_j, j=s,p$ of the incident field onto the local $\boldsymbol{e}_s$ and $\boldsymbol{e}_p$ direction (see Fig.~\ref{fig1}e) multiplied by the appropriate reflection coefficient.
$\mathbf{R}$ acts on an initial polarization $\boldsymbol{E}$, which, for the central wave vector, is a 2D constant Jones vector with components $E_x$ and $E_y$.
This is the global polarization of the incident field which we distinguish from the local polarization in terms of $s$ and $p$ components.
Because of transversality, however, the other plane waves in the spectrum have a small $z$ component such that $\boldsymbol{E} = (E_x, E_y, -[E_x \cos \alpha + E_y \sin \alpha] \tan \delta).$ 
As we only consider a polarized component of the beam, the reflected beam is filtered by a constant polarization analyser $\boldsymbol{F}$ with components $(F_x, F_y, 0).$
All of the physics of beam reflections can be explained by a Taylor expansion about $\delta = 0$ of the scalar multiplication operator $\boldsymbol{F}^*\cdot\mathbf{R}\cdot\boldsymbol{E}$ in Fourier space as we briefly summarize in the following (also see \cite{Bliokh+:OL34:2009}).

In the transverse plane the real-space scalar reflected beam is given by 
\begin{equation}
   \psi(\boldsymbol{r}) 
   = \int_0 d\delta \int_{-\pi}^\pi d\alpha \sigma(\delta) \delta \mathrm{e}^{\mathrm{i} k \sin \delta \boldsymbol{r} \cdot (\cos \alpha, \sin \alpha) - \mathrm{i} \ell \alpha} \boldsymbol{F}^* \cdot \mathbf{R} \cdot \boldsymbol{E},
   \label{eq:reflected}
\end{equation}
where $\boldsymbol{r}=(x,y)$ and $k$ is the wavenumber of the incident light.
The formula for the beam shift follows from a first order Taylor expansion of the reflection matrix term,
\begin{equation}
   \boldsymbol{F}^* \cdot \mathbf{R} \cdot \boldsymbol{E} \approx \overline{R} \left[ 1+ \delta \boldsymbol{\mathcal{D}} \cdot (\cos \alpha, \sin \alpha) \right] 
   \approx \overline{R} \mathrm{e}^{ \delta \boldsymbol{\mathcal{D}} \cdot (\cos \alpha, \sin \alpha)},
   \label{eq:refmatrix}
\end{equation}
where $\boldsymbol{\mathcal{D}} =(\mathcal{D}_x, \mathcal{D}_y)$  is a complex 2-vector with components $\mathcal{D}_x = (F^\ast_x E_x r_p' - F_y^\ast E_y r_s')/(F^\ast_x E_x r_p - F_y^\ast E_y r_s)$ and $\mathcal{D}_y = (F^\ast_y E_x + F^\ast_x E_y) \cot \theta_0 (r_p+r_s)/(F^\ast_x E_x r_p - F_y^\ast E_y r_s).$

When $\ell = 0$, the spatial shift is given by $-\mathrm{Im} \boldsymbol{\mathcal{D}}/k,$ which adds to the $(x,y)$-dependent term in the exponent in (\ref{eq:reflected}) when identifying $\sin \delta$ with $\delta$.
This is the well-known Artmann formula for spatial beam shifts \cite{Artmann:AndP437:1948}.
The angular shift is the angular mean of the Fourier transform of $\psi$, which is given by $\sigma_2 \mathrm{Re} \boldsymbol{\mathcal{D}}$ \cite{GoetteDennis:NJP:2012,*[{An alternative, but equivalent definition of the angular shift can be found in }] [] AielloWoerdman:OL33:2008}.
For $\ell \neq 0$ the net vortex beam shift is 
\begin{equation}
   \boldsymbol{D}_{\mathrm{centroid}} = -\mathrm{Im} \boldsymbol{\mathcal{D}} - \mathrm{i} \ell \boldsymbol{\sigma}_2 \cdot \mathrm{Re} \boldsymbol{\mathcal{D}},
   \label{eq:vortexinduced}
\end{equation}
where $\boldsymbol{\sigma}_2$ is the second Pauli matrix \cite{Merano+:PRA82:2010,*[] [{, in which a different sign convention for $\ell$ is used.}] DennisGoette:NJP:2012}. 
The first term here is the usual Artmann translation; the second term, usually associated with the angular shift, comes from the azimuthal complex amplitude structure of the vortex beam.

\begin{figure}
\begin{center}
\includegraphics[width=0.49\textwidth]{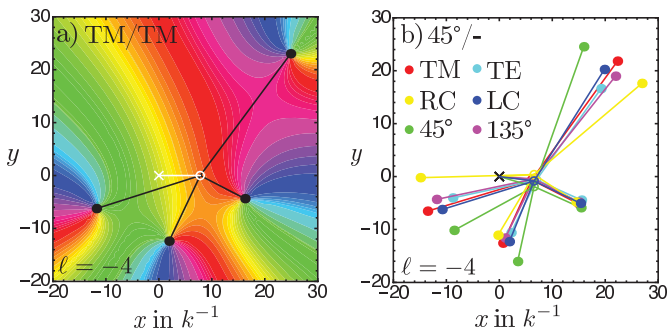}
\caption{\label{fig2}   (Colour online)
   a) Plot comparing the constellation of vortices obtained from the roots of the local expansion in (\ref{eq:localexp}) with a numerical calculation of the phase of the field.
   The incident field is a Bessel beam incident at $\theta_0 = 46^\circ$ with $\ell = -4$ with a fixed opening angle of $\delta=0.01$. 
   The incident polarization $\boldsymbol{E}$ and the analyser $\boldsymbol{F}$ are both oriented in the $x$ direction (TM/TM). 
   b) Plot showing the variation of constellations for $45^\circ$ diagonal incident polarisation and different analyser settings including linear in the $y$ direction (TE), right (RC) and left (LC) circular polarization, as well as $45^\circ$ and $135^\circ$ diagonal polarization.
 }
\end{center}
\end{figure}

The shift to the intensity centroid of a vortex-carrying beam is different from the shift of the vortex itself \cite{OkudaSasada:OE14:2006,BliokhDesyatnikov:PRA79:2009}. 
The translation $-\mathrm{Im} \boldsymbol{\mathcal{D}}$ affects the entire beam including vortex and the centroid.
However, $\mathrm{Re} \boldsymbol{\mathcal{D}},$ being associated with a change in the profile of the beam, affects them differently: for an asymmetric profile, a vortex, being an absence of intensity, repels the intensity centroid, and it follows form the more general argument below that the shift of a vortex in a $|\ell| = 1$ beam is
\begin{equation}
   \boldsymbol{D}_\mathrm{vortex} = -\mathrm{Im} \boldsymbol{\mathcal{D}} \pm \mathrm{i} \boldsymbol{\sigma}_2 \cdot \mathrm{Re} \boldsymbol{\mathcal{D}},
   \label{eq:vortexmean}
\end{equation}
where $\pm$ refers to $\ell \gtrless 0$.

When $|\ell| > 1,$ topology preserves the overall vortex charge, but the symmetry is broken and typically there is a constellation of $|\ell|$ unit-charge vortices in the reflected beam, with separation comparable to the beam shift itself.
An example of a numerical calculation of the reflection of a $\ell = -4$ Bessel beam \cite{McGloinDholakia:JMO46:2005}, with several different incident and analyzer polarizations, is shown in Fig.~\ref{fig2}; the constellation is not simply a regular polygon or `row' \cite{Dennis:OL31:2006}, but a more complicated function of incident and analyzer polarization, incidence angle $\theta_0$ and refractive index $n.$
The centroid of the vortices, represented by the white circle, is the same as the position of the single shifted vortex $\boldsymbol{D}_{\mathrm{ vortex}}$: the centroid of the $|\ell|$ vortex points is a discrete, topological counterpart to the shift of the intensity centroid, but without the $\ell$ weighting, anticipated in \cite{BliokhDesyatnikov:PRA79:2009}.

To analyse these constellations we now derive an analytic approximation for the reflected vortex beam close to the beam axis, as a complex polynomial in $x + i y$ or $x - i y$ depending on the sign of $\ell$ \cite{Dennis:OL31:2006}, by collecting all the terms of the lowest order in $\delta$ in the expansion of the reflection matrix (\ref{eq:refmatrix}).
Up to order $|\ell|,$ this can be written as
\begin{equation}
   \boldsymbol{F}^* \cdot \mathbf{R} \cdot \boldsymbol{E} \approx \overline{R} \left( 1 + \delta C_1 + \frac{1}{2} \delta^2 C_2 + \dots \frac{1}{|\ell|!} \delta^{|\ell|} C_{|\ell|} \right),
   \label{eq:reflectionexpansion}
\end{equation}
where $\overline{R} = \boldsymbol{F}^* \cdot \mathbf{R} \cdot \boldsymbol{E} \vert_{\delta = 0}$. 
The original beam shift follows from $C_1 = \boldsymbol{\mathcal{D}} \cdot (\cos \alpha, \sin \alpha),$ which contains only single powers of $\sin \alpha$ and $\cos \alpha$ and depends on the first derivatives of the reflection coefficients.
The higher coefficients $C_n$ contain combinations of $\sin^u \alpha \cos^v \alpha$ and $r^{(q)}_j = (d^q/d \delta^q) r_j \vert_{\delta = 0}$ for $j=s,p$ with $u+v =n$ and $q \le u,v.$
Each $C_n$ may thus be written as a complex Fourier series
\begin{equation}
   C_n = \frac{c_n^+ \mathrm{e}^{\mathrm{i} n \alpha} + c_n^- \mathrm{e}^{-\mathrm{i} n \alpha}}{2^n} + \dots.
\end{equation} 
This expansion of the reflection matrix is analogous to an aberration expansion in terms of complex Zernike modes \cite{Zernike:Physica1:1934}, with $n$ giving the order of aberration: $c_1^\pm$ corresponds to tilt, $c^{\pm}_2$ to astigmatism, etc.

The core of the reflected vortex beam can be found by combining Eqs.~(\ref{eq:reflected}) and (\ref{eq:reflectionexpansion}); after integration over the azimuthal $\alpha,$ each $\mathrm{e}^{\pm \mathrm{i} q \alpha}$ contributes a Bessel function $J_{\ell \mp q},$ which is Taylor approximated close to the axis.
The lowest order term in the beam is $\delta^{|\ell|},$ which comes from each $c_n^{\pm}$ for $\ell \gtrless 0;$ all other terms have higher order and therefore smaller contribution.
The size of the $n$th contribution, corresponding to $c^{\pm}_n$ is
\begin{multline}
   \frac{c_n^\pm \delta^n}{2^n n!} \int_{-\pi}^\pi d \alpha \mathrm{e}^{\mathrm{i} k \delta r \cos(\alpha - \phi) - \mathrm{i} (\ell \mp  n) \alpha} 
   \\ \; \approx \delta^{|\ell|} c_n^\pm 2\pi \mathrm{i}^{|\ell|-n}[2^{|\ell|} n! (|\ell|-n)!]^{-1} \left[k (x \mp \mathrm{i} y)\right]^{|\ell|-n}. 
\end{multline}
substituting $r \exp(\mp \mathrm{i} \phi) = x \mp \mathrm{i} y.$

On collecting common factors, the reflected field near the axis for a given $\ell$ is proportional to
\begin{equation}
   \psi \propto \zeta^{|\ell|} - \mathrm{i} |\ell| c_1^\pm \zeta^{|\ell|-1} - \frac{|\ell|^2 - |\ell|}{2} c_2^\pm \zeta^{|\ell|-2}  + \dots + (-\mathrm{i})^{|\ell|} c_{|\ell|}^\pm, 
   \label{eq:localexp}
\end{equation}
where we have used $\zeta = k(x \mp \mathrm{i} y)$ with `$-$' for $\ell > 0$ and `$+$' for $\ell<0$.
Each root of this complex polynomial corresponds to a vortex, and the constellation depends on the roots of the polynomial with coefficients given by $(-\mathrm{i})^n c_n^\pm$ times a binomial coefficient as in (\ref{eq:localexp}).
For increasing $|\ell|$ the polynomials in (\ref{eq:localexp}) form an Appell sequence \cite{Roman:Dover:2005}, with coefficients generated by  the sequence $c_n^{±}$ of coefficients from the Zernike expansion of the reflection matrix, for which there are many relations between the roots and the coefficients \cite{Marden:AMS:1966}.
The integral over $\delta$ as in (\ref{eq:reflected}) does not affect Eq.~(\ref{eq:localexp}), since these terms have the same, lowest-order contribution in $\delta,$ and so the constellation is independent of the spectrum and hence the radial profile of the beam, as long as it is narrowly confined in Fourier space.

In the complex plane, the arithmetic mean (centroid) of the roots of a polynomial of the form (\ref{eq:localexp}) is given by $\mathrm{i} c_1^\pm$  \cite{Marden:AMS:1966}, which proves that the mean position of the vortices is always given by $\mathrm{i} c_1^\pm = \boldsymbol{\mathcal{D}} \cdot (\mathrm{i},\pm \mathrm{1})$.
The vortex centroid shift is independent of $\ell,$ and can be understood purely in terms of the representation of the beam as a polynomial in the complex $(x\mp \mathrm{i} y)$ plane.

The vortex constellation for $|\ell| >1$ is sensitive to terms beyond the first-order shift.
The second order term $c_2^{\pm}$ is related to the complex astigmatism of $\boldsymbol{F}^*\cdot\mathbf{R}\cdot\boldsymbol{E},$ proportional to $\mathrm{e}^{\pm 2 i \alpha}.$
When $|\ell| = 2,$ Eq.~(\ref{eq:localexp}) is a quadratic equation with $c_2^{\pm}$ the `constant'; $\mathrm{i} c^{\pm}_1$ is the midpoint of the two roots, and their separation from the midpoint is $\pm\sqrt{c_2^{\pm}-(c_1^{\pm})^2}$ (where the complex argument represents the direction).
Measurement of the two singularity positions would therefore provide the two complex numbers $c_1^{\pm}$ and $c_2^{\pm},$ from the first two orders of the Zernike expansion of the reflection coefficient. The former contains zeroth and first derivatives of the reflection coefficients via the complex shift vector $\boldsymbol{\mathcal{D}}$ and the latter is given by $c_2^\pm = (F_x^\ast, F_y^\ast) \cdot \boldsymbol{\mathfrak{c}}_2^\pm \cdot  (E_x, E_y)/(F^\ast_x E_x r_p - F_y^\ast E_y r_s),$ where the $\boldsymbol{\mathfrak{c}}_2^\pm$ matrix is given by
\begin{multline}
\boldsymbol{\mathfrak{c}}_2^\pm = (r_s + r_p) \cot^2 \theta_0 \begin{pmatrix} 1 & \pm \mathrm{i} \\ \pm \mathrm{i} \, \mathrm{sc}^2 & -1  \end{pmatrix} 
+ \frac{1}{2} \cot \theta_0 \begin{pmatrix} -r_p' & 0 \\ 0 & r_s' \end{pmatrix} \\
\mp \mathrm{i} (r_s'+r_p') \cot \theta_0 \begin{pmatrix} 0 & 1 \\ 1 & 0 \end{pmatrix} + \frac{1}{2}  \begin{pmatrix} r_p'' & 0 \\ 0 & -r_s'' \end{pmatrix},
\end{multline}
where $\mathrm{sc}^2 = \sec^2 \theta_0$.
This complicated form of $\boldsymbol{\mathfrak{c}}_2^\pm$ shows that for higher order shift effects a separation in terms of diffractive corrections and optical spin-orbit interaction is no longer possible \cite{BliokhAiello:CUP:2012}.

\begin{figure}
\begin{center}
\includegraphics[width=0.49\textwidth]{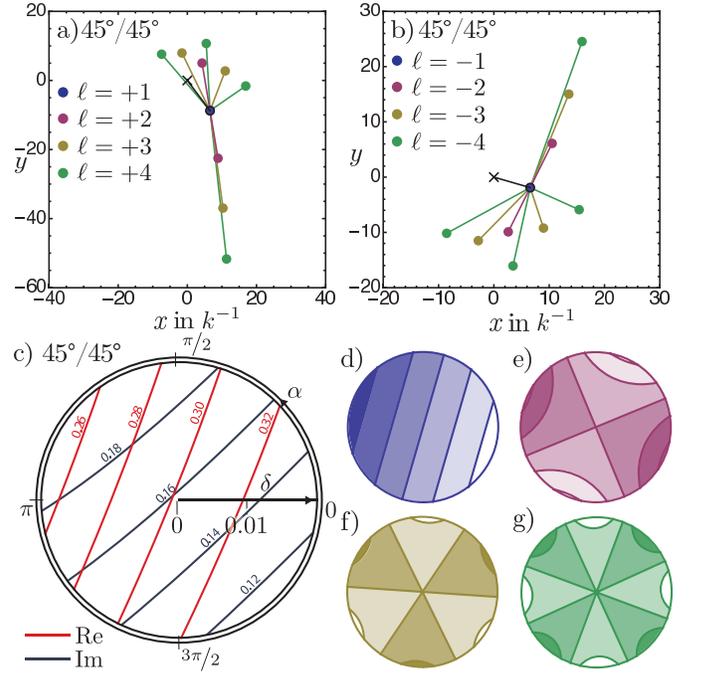}
\end{center}
\caption{\label{fig:3} (Colour online)
Vortex constellations for increasing positive $\ell$ (a) and negative $\ell$ (b). 
c) Contours in Fourier space (conical beam geometry) of the real and imaginary part of $ \boldsymbol{F}^* \cdot \mathbf{R} \cdot \boldsymbol{E}$ for $\theta_0 = 46^\circ$ and both $\boldsymbol{F}$ and $\boldsymbol{E}$ set to $45^\circ$ diagonal polarization. 
d-g) Contours plots for of the real parts of the coefficients in (\ref{eq:localexp}) multiplied by the appropriate Fourier factor $\exp(-\mathrm{i} n \alpha)$ for comparison with $\ell<0$  in b). d) $n = 1,$ e)  $n = 2,$ f)  $n= 3,$ g)  $n = 4.$ (Scaling identical to figure c). The gradient over the contours  in d) corresponds to the shift as indicated by the black line between $\times$ and $\circ$ in b).}  
\end{figure}

In Fig.~\ref{fig:3} we illustrate how this `singularimetry' would give increasing information as positive $\ell$ increases (a) and as negative $\ell$ increases (b).
As $|\ell|$ increases, the constellations increase in complexity, but maintaining common features (such as their centre), as their coefficients are algebraically related in Eq.~(\ref{eq:localexp}). 
These complex coefficients are proportional to the $c^{\pm}_{n},$ which correspond to succesive azimuthal aberration terms (tilt, astigmatism, ...).
This is shown in Fig.~\ref{fig:3} c) - g), which illustrates how a fixed reflection matrix term $\boldsymbol{F}^*\cdot\mathbf{R}\cdot\boldsymbol{E}$ as a function of $\alpha,\delta$ (c) is decomposed into aberration terms of increasing order. 

It has recently been appreciated that there is a close connection between beam shifts and the notion of quantum weak values \cite{ *[{see }] [{ and references therein.}] AharonovRohrlich:WVCH:2005}.
We note that going beyond $|\ell|=1$ shows that there is weak information beyond the first order weak value.

Our argument extends beyond reflection, to any paraxial vortex beam experiencing an aberration-like multiplication in Fourier space: rather than a single order $\ell$ on-axis vortex, the beam will have a constellation of $\ell$ vortices which can be represented by a polynomial with coefficients dependent on the azimuthal terms. 
This effect of topological aberration will affect any vortex beam undergoing oblique reflection, refraction in a real optical system, suggesting a fundamental limitation on the purity of quantized OAM modes in optical devices \cite{Agnew+:PRA84:2011, *Loeffler+:arxiv:2012}.
However, as the connection between vortex positions and the aberrations is so algebraically direct, any singularimetric device which measures the precise vortex distribution has direct access to the beam shift and later terms, without any further analysis of the overall beam.

\begin{acknowledgments}
We would like to thank Michael Berry, Konstantin Bliokh and James Ring for careful reading and helpful comments. MRD and JBG acknowledge financial support from the Royal Society as University Research Fellow (MRD) and Newton Alumnus (JBG).
\end{acknowledgments}


%

\end{document}